\documentclass[twocolumn,aps,showpacs]{revtex4}
\usepackage{graphicx}

\begin{document}

\title{In~situ x-ray diffraction study of epitaxial growth of ordered Fe$_3$Si films}
\author{B.~Jenichen}\email { jen@pdi-berlin.de}
\author{V. M.~Kaganer}
\author{W.~Braun}
\author{R.~Shayduk}
\author{B.~Tinkham}
\author{J.~Herfort}

\affiliation{Paul-Drude-Institut f\"{u}r Festk\"{o}rperelektronik,
Hausvogteiplatz 5--7, D-10117 Berlin, Germany}
\date{\today}

\begin{abstract}
Molecular beam epitaxy of Fe$_3$Si on GaAs(001) is studied in situ
by grazing incidence x-ray diffraction. Layer-by-layer growth of
Fe$_3$Si films is observed at a low growth rate and substrate
temperatures near 200\thinspace $^{\circ }$C. A damping of x-ray
intensity oscillations due to a gradual surface roughening during
growth is found. The corresponding sequence of coverages of the
different terrace levels is obtained. The after-deposition surface
recovery is very slow. Annealing at 310\thinspace $^{\circ }$C
combined with the deposition of one monolayer of Fe$_3$Si restores
the surface to high perfection and minimal roughness. Our
stoichiometric films possess long-range order and a high quality
heteroepitaxial interface.
\end{abstract}

\pacs{81.15.Hi, 61.10.Nz, 68.35.Bs, 75.50 Cc}

\maketitle



Combinations of magnetic and semiconducting materials lead to
further development in the field of magnetoelectronics
\cite{prinz90, parkin04}. Ferromagnet/semiconductor
heterostructures should possess a rather perfect interface
\cite{kratzer05} to minimize the scattering of spins. High thermal
stability is required for device processing and operation.
Fe$_{3}$Si on GaAs is a promising candidate for magnetoelectronic
applications \cite{hong91, liou93,ionescu05}. The Curie
temperature of Fe$_{3}$Si is as high as  567\thinspace $^{\circ
}$C \cite{nakamura88} and the Fe$_{3}$Si / GaAs interface is
stable up to more than 400\thinspace $^{\circ }$C
\cite{herfort05}. Spin injection through the Fe$_{3}$Si / GaAs
interface at room temperature has been demonstrated recently
\cite{kawa04}. The lattice misfit between stoichiometric
Fe$_{3}$Si and GaAs is very small \cite{herfort03,herfort06} and
allows for the formation of a perfect interface without strain
releasing defects like misfit dislocations. The lattice match and
the long range order of the as-grown thin epitaxial films depend
on the stoichiometry \cite{jen05}. Significant changes of the
saturation magnetization and the sheet resistance with film
stoichiometry have been observed \cite{herfort04}.

The layer-by-layer growth mode allows for well controlled
fabrication of epitaxial layer sequences with very sharp
interfaces between them. The aim of the present work is the
\textit{in situ} characterization of the Fe$_{3}$Si epitaxial
growth process in the layer-by-layer growth mode by x-ray surface
diffraction methods. The surface and interface roughness and long
range order in the films are monitored under ultrahigh vacuum
conditions. We use grazing incidence x-ray diffraction (GID) which
is ideally suited for thin film investigations. Fe$_{3}$Si films
can be grown by molecular beam epitaxy (MBE) at GaAs substrate
temperatures near 200\thinspace $^{\circ }$C \cite{herfort03}.
They were grown for the present study in an MBE chamber inserted
onto the diffractometer at the wiggler beamline U125/2 KMC
\cite{jeni03} at the storage ring BESSY in Berlin. A double
crystal Si(111) monochromator was used. The energy of the
radiation was 10~keV and the incidence angle was kept at
0.3\thinspace $^{\circ }$. The acceptance angle of the detector
was 0.1\thinspace $^{\circ }$ both perpendicular  and parallel to
the surface.

\begin{figure}[b]
\includegraphics[width=7cm]{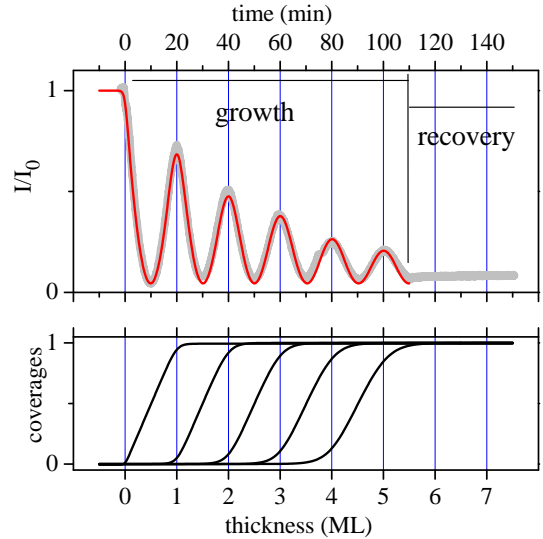}
\caption{(a) (Color online) X-ray intensity oscillations (gray
line) observed by GID at the 1~1~0.05 reflection during MBE
layer-by-layer growth of an Fe$_3$Si epitaxial film on a several
nm thick Fe$_3$Si buffer layer on a GaAs(001) substrate. The
growth temperature of the film and the buffer was 220$^{\circ }$C.
The oscillations are damped and the recovery of the x-ray
intensity after deposition is slow. Fit (fine red line) of the
interference function (\ref{eq1}) to the experimental curve. (b)
Coverages obtained by this fitting.
   } \label{fig:Fig1}
\end{figure}

GaAs(001) templates were prepared in a separate III-V growth
chamber using standard GaAs growth techniques. The sample was then
capped by As and transferred into the MBE system at BESSY for the
Fe$_{3}$Si deposition by means of an ultrahigh vacuum shuttle. The
As cap was removed by annealing the sample in the preparation
chamber at a temperature of 350$^{\circ }$C before transferring it
into the growth chamber. The Fe$_{3}$Si layers were then grown on
the As-rich c($4\times4$) reconstructed GaAs surface at different
substrate temperatures near 200$^{\circ }$C with a growth rate of
3 monolayers (ML) per hour, similar to the procedure described in
Ref.~\onlinecite{herfort03} (1 ML $= 0.5 a \cong 0.28~$nm, where
$a$ is the lattice parameter). The Si and Fe cell temperatures
were tuned in order to obtain a perfect lattice match of the
films. The position of the Fe$_{3}$Si layer peak of the x-ray
diffraction curve was monitored in order to reach a coincidence
with the corresponding GaAs peak. We obtained optimum temperatures
of 1239$^{\circ }$C and 1370$^{\circ }$C for the Fe and the Si
cells, respectively.

\begin{figure}[t]
\includegraphics[width=7cm]{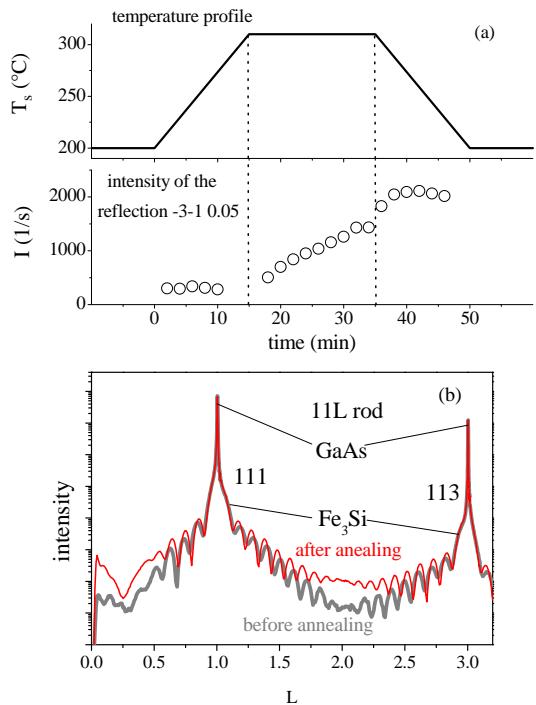}
\caption{(Color online) Annealing of the Fe$_3$Si epitaxial layer.
(a) Temperature profile together with the measured intensity of
the surface sensitive reflection $\bar{3}~\bar{1}~0.05$. The
intensity rises during annealing by an order of magnitude. The
whole $1~1~L$ rod is shown in (b). The positions of the maxima of
the narrow GaAs peaks coincide with the centers of the broader
Fe$_3$Si peaks indicating lattice matching. The maximum near
$1~1~0.05$ increases by  more than a factor of ten thanks to
annealing and the decay of the crystal truncation rod between the
bulk reflections is considerably reduced because of lower surface
roughness.} \label{fig:Fig2}
\end{figure}

\begin{figure*}[h!]
\includegraphics[width=\textwidth]{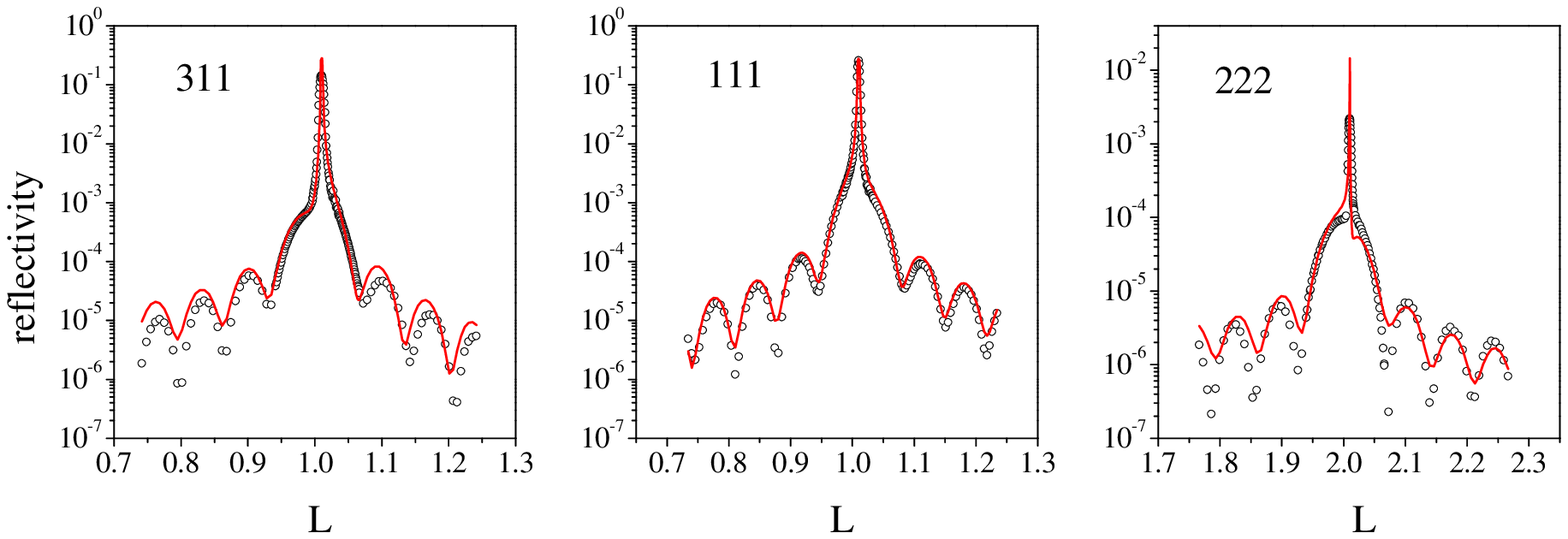}
\caption{(Color online) Comparison of the experimental CTRs
(circles) and the corresponding simulations (red lines) assuming a
fully ordered 29~ML thick Fe$_3$Si lattice. The Fe$_3$Si peaks are
superlattice reflections sensitive to disorder. The film was grown
at a substrate temperature of 200$^{\circ }$C. The maxima of the
narrow GaAs peaks coincide with the centers of the broader
Fe$_3$Si peaks indicating lattice matching. The intensities of
Fe$_3$Si peaks reach the theoretical values, supporting the
assumption of a fully ordered film and indicating a stoichiometric
composition of the film.} \label{fig:Fig3}
\end{figure*}

Layer-by-layer growth is observed by x-ray intensity oscillations
during deposition of Fe$_{3}$Si, Fig.~1~(a). The x-ray intensity
is measured in the reciprocal space point 1~1~0.05, close the bulk
forbidden reflection 1~1~0. The oscillations are continuously
damped, indicating gradual surface roughening. The oscillation
period $T$ is $1220\pm10$~s. Such a low growth rate facilitates
the formation of a highly ordered Fe$_{3}$Si film. Recovery of the
diffracted intensity after growth is very slow. In order to
overcome this slow recovery and restore the initial flat surface,
the surface was annealed at 310$^{\circ }$C for about 20~min with
simultaneous deposition of one monolayer of Fe$_{3}$Si. This
procedure considerably improves the flatness of the Fe$_{3}$Si
surface, as observed by the increase of the intensity of bulk
forbidden surface sensitive reflections by a factor of about ten,
see Fig.~2~(a). The same annealing procedure was also used for
surface preparation before the measurement shown in Fig.~1~(a).

Crystal truncation rods (CTRs) presented in Fig.~2~(b) demonstrate
the results of growth and the effect of annealing in more detail.
A nearly perfect angular coincidence of the Fe$_{3}$Si and GaAs
peaks is achieved by the fine tuning of the Fe and Si cell
temperatures. An almost perfect lattice match of the films during
and after growth is attained. The Fe$_{3}$Si peaks are broader and
less intense due to small thickness of the film. The shape of the
CTRs in the middle between bulk diffraction peaks is a measure of
the surface roughness \cite{robinson86}. High intensity indicates
a well-ordered, highly planar surface. The intensity of the 1~1~L
rod in Fig.~2~(b) in the vicinity of the bulk forbidden
reflections 1~1~0 and 1~1~2 increase by an order of magnitude
during annealing. Thus, the original high quality of the surface
can be restored by the annealing cycle. Pronounced thickness
fringes observed on the CTR imply smooth and parallel top and
bottom interfaces. These fringes evidence not only a smooth
surface but also a a high quality of the Fe$_{3}$Si / GaAs
interface \cite{jen07}. We checked this by means of simulation of
the corresponding x-ray reflectivity curves.

The kinematical diffraction theory from stepped surfaces
\cite{cohen89} describes the average surface shape during growth
in terms of coverages $\Theta_n(t)$ of successive levels. In the
ideal case, growth starts (at $t=0$) from a perfectly flat surface
with $\Theta_0 = 1$ and $\Theta_n = 0$ for $n\geq1$ ($n$ is
integer). During growth, the coverages run from 0 to 1 with
$\Theta_{n+1}<\Theta_n$, Fig.~1 (b). Two successive terrace levels
interfere completely destructively under the antiphase diffraction
condition. For the zinc blende structure, this condition is met
for reflections $HKL$ with odd $H,K$ at $L=0$, i.e., for
simultaneously grazing incidence and grazing exit diffraction
conditions \cite{braun03}. The destructive interference at $L=0$
originates from the lateral shift of the successive crystalline
layers in the zinc blende structure, which accompanies the
vertical shift.

The diffracted intensity at the antiphase condition is proportional
to the interference function
\begin{equation}
S(q)=|F_{hkL}|^2|C_0-C_1+C_2-...|^2,
\label{eq1}
\end{equation}
where $F_{hkL}$ is the structure amplitude of the given reflection
and $C_n = \Theta_n-\Theta_{n+1}$ is the exposed coverage on the
$n$th level of the surface \cite{cohen89}. The time dependence of
the layer coverage can be described by the continuity equation
\begin{equation}
d\Theta_n/dt=(J_{n-1}-J_{n})/T, \label{eq2}
\end{equation}
where $T$ is the time required for the deposition of one monolayer.
For ideal layer-by-layer growth, the fluxes $J_n(t)$ are the
Heaviside step functions, $J_n=H(t-nT)$. An imperfect layer-by-layer
growth can be described by the smoothed functions \cite{braun03}
\begin{equation}
J_n(t)= \mathrm{tanh}[(t-nT)/b_n]. \label{eq3}
\end{equation}
The changes in the onset and completion of the formation of
subsequent monolayers can be modeled by choosing appropriate
widths $b_n$. We assume according to Ref. \cite{braun03}, a power
law dependence, $b_n={\gamma}n^{\delta}$. The layer coverages
$\Theta_n(t)$ shown in Fig.~1~(b) are obtained from a fit to
measured growth oscillations with the parameters $\gamma=0.29$ and
$\delta=0.7$. Figure~1~(b) presents the resulting time dependence
of the coverages.

Intensities of the Fe$_3$Si reflections depend on the long-range
order of the atoms in the sublattices. Fe$_{3}$Si has the face
centered cubic D0$_{3}$ structure that can be viewed as a fcc
lattice with the basis consisting of four atoms A, B, C, and D
with coordinates A(0, 0, 0), B(0.25, 0.25, 0.25), C(0.5, 0.5,
0.5), and D(0.75, 0.75, 0.75) \cite{niculescu76, hines76, jen05}.
In the ordered Fe$_{3}$Si crystal, Fe atoms occupy three
sublattices A, B, and C, while Si atoms take the sublattice D. The
disorder is described by two order parameters $\alpha $ and $\beta
$, which are fractions of Si atoms occupying the Fe(B) and Fe(A,C)
sites, respectively. The reflections from a partially disordered
crystal can be divided into three types, fundamental reflections
and two types of superlattice reflections. Fundamental
reflections, that are not influenced by disorder, are given by
$H+K+L=4n$, where $n$ is an integer. The structure amplitude of
such reflections is
\begin{equation}
F_{0}=4(f_{\mathrm{Si}}+3f_{\mathrm{Fe}}),  \label{eq4}
\end{equation}
where $f_{\mathrm{Si}}$ and $f_{\mathrm{Fe}}$ are atomic scattering
factors of the respective elements. Reflections with odd $H,K,L$ are
sensitive to both types of disorder, the structure amplitude being
\begin{equation}
F_{1}=4i(1-2\alpha -\beta )(f_{\mathrm{Si}}-f_{\mathrm{Fe}}).
\label{eq5}
\end{equation}
Reflections given by $H+K+L=4n-2$ are sensitive to disorder in the
Fe(A,C) sublattice,
\begin{equation}
F_{2}=-4(1-2\beta )(f_{\mathrm{Si}}-f_{\mathrm{Fe}}). \label{eq6}
\end{equation}

Figure 3 compares measured and calculated x-ray diffraction curves
for the reflections that are sensitive to different types of
disorder: 1~1~1 and 3~1~1 reflections are sensitive to disorder in
both sublattices, see Eq.\ \ref{eq5}, while the 2~2~2 reflection
is sensitive to disorder in the Fe(A,C) sublattice, see Eq.\
\ref{eq6}. The x-ray intensity distribution along each CTR was
calculated as a sum of the amplitudes for two-beam dynamical
diffraction problems corresponding to different reciprocal lattice
points along the given CTR \cite{kaganer07}. 16 reflections were
included in the sum. The Fe$_3$Si film thickness obtained from the
intensity calculations is 29 ML. The calculations are performed
for a fully ordered film, $\alpha=\beta=0$. The experimental
curves were scaled to the calculated ones by requiring equal
integrated intensities of the substrate peaks. With this scaling,
the intensities of the Fe$_3$Si peaks reach the theoretical
values, thus proving the full order of the film. Similar results
were obtained for other lattice matched films grown at low growth
rate.
In conclusion, we have found the layer-by-layer growth conditions
during MBE of Fe$_{3}$Si films on GaAs (001) substrates at low
growth rates ($\approx$~3~ML/h) and performed \textit{in situ}
experiments using grazing incidence x-ray diffraction. An
excellent surface and interface quality was achieved.  The growth
of lattice matched films at such a low rate leads to a high degree
of long-range order.
The authors thank Claudia Herrmann, Steffen Behnke, Hans-Peter
Sch\"onherr and Tatsuro Watahiki for support during the
experiments, and Uwe Jahn and Achim Trampert for critical reading
of the manuscript.


\end{document}